\documentclass[twocolumn,superscriptaddress,prl,preprintnumbers,amsmath,amssymb%
,floatfix,showpacs]{revtex4}

\usepackage{graphicx}
\begin{document}

\title{Penetration depth, symmetry breaking, and gap nodes\\ in superconducting $\bf PrOs_4Sb_{12}$}

\author{Lei Shu}
\author{D. E. MacLaughlin}

\affiliation{Department of Physics, University of California, Riverside, California 92521-0413}

\author{R. H. Heffner}
\affiliation{MS K764, Los Alamos National Laboratory, Los Alamos, New Mexico 87545}

\author{G. D. Morris}
\altaffiliation[Permanent address: ]{TRIUMF, Vancouver, Canada \mbox{V6T 2A3}.}
\affiliation{MS K764, Los Alamos National Laboratory, Los Alamos, New Mexico 87545}

\author{O. O. Bernal}
\affiliation{Department of Physics and Astronomy, California State University, \\Los Angeles, California 90032}

\author{F. Callaghan}
\affiliation{Department of Physics, Simon Fraser University, Burnaby, B.C., Canada \mbox{V5A 1S6}}

\author{J. E. Sonier}
\affiliation{Department of Physics, Simon Fraser University, Burnaby, B.C., Canada \mbox{V5A 1S6}}
\affiliation{Canadian Institute for Advanced Research, Toronto, Ontario, Canada \mbox{M5G 1Z8}}

\author{A. Bosse}
\affiliation{Department of Physics, University of California, Riverside, California 92521-0413}
\affiliation{Inst.\ f.\ Metallphysik u.\ Nukleare Festk\"orperphysik, Technische Univ.\ Braunschweig, 38106 Braunschweig, Germany}

\author{J. E. Anderson}
\affiliation{Department of Physics, University of California, Riverside, California 92521-0413}

\author{W. M. Yuhasz}
\author{N. A. Frederick}
\author{M. B. Maple}
\affiliation{Department of Physics, University of California, San Diego,\\ La Jolla, California 92093-0319}

\date{\today}

\begin{abstract}
Transverse-field muon spin relaxation rates in single crystals of the heavy-fermion superconductor PrOs$_4$Sb$_{12}$ ($T_c = 1.85$~K) are nearly constant in the vortex state for temperatures below ${\sim}0.5T_c$. This suggests that the superconducting penetration depth~$\lambda(T)$ is temperature-independent at low temperatures, consistent with a gapped quasiparticle excitation spectrum. In contrast, radiofrequency measurements yield a stronger temperature dependence of $\lambda(T)$, indicative of point nodes in the gap. A similar discrepancy exists in superconducting Sr$_2$RuO$_4$ which, like PrOs$_4$Sb$_{12}$, breaks time-reversal symmetry (TRS) below $T_c$, but not in a number of non-TRS-breaking superconductors.
\end{abstract}

\maketitle

Phase transitions are always symmetry breaking; the symmetry of the ordered state is less than the full symmetry of the Hamiltonian. In all superconductors gauge symmetry is broken and the ground-state wave function acquires a definite phase. Additional symmetries are simultaneously broken in so-called unconventional superconductors. Examples include the point-group symmetry of the lattice, spin rotation symmetry, and time-reversal symmetry (TRS)~\cite{SiUe91,HeNo96}. TRS is always broken in magnetic transitions but rarely in superconductors. Probably the most direct evidence for TRS breaking is the onset of a spontaneous local magnetic field in zero applied field below the superconducting transition temperature~$T_c$; the local field reverses its direction under time reversal and thus signals broken TRS\@. 

Muon spin rotation/relaxation ($\mu$SR) experiments have proved invaluable in characterizing the superconducting state~\cite{SBK00}. In the $\mu$SR technique~\cite{Brew94} spin-polarized positive muons ($\mu^+$) are implanted in the sample and precess in the local magnetic field. This precession is detected using the asymmetry of the $\mu^+$ beta decay (the decay positron is emitted preferentially in the direction of the $\mu^+$ spin). The distribution of $\mu^+$ precession frequencies directly reflects the distribution of magnetic fields in the sample. Reproducible $\mu$SR evidence for a TRS-breaking local field has been found in only three superconductors: the heavy-fermion system~(U,Th)Be$_{13}$~\cite{HSWB90}, the transition-metal oxide~Sr$_2$RuO$_4$~\cite{LFKL98}, and the Pr-based heavy-fermion compound~PrOs$_4$Sb$_{12}$~\cite{ATKS03}.

TRS breaking in a superconductor may (but need not) be accompanied by breaking of other symmetries~\cite{SiUe91}. The most accessible consequence of additional symmetry breaking is the presence of point or line nodes $\Delta({\bf k}) = 0$ in the superconducting energy gap function~$\Delta({\bf k})$ on the Fermi surface. Thermal excitations of low-lying states are strongly modified by gap nodes, and these excitations govern superconducting properties at low temperatures. The temperature dependence of the superconducting penetration depth~$\lambda$, in particular, depends on details of the gap structure~\cite{SiUe91,HeNo96}: for a nodeless superconductor $\Delta\lambda(T) = \lambda(T) - \lambda(0) \propto e^{-\Delta/T}$, which is small at low temperatures, whereas with gap nodes $\Delta\lambda \propto T^n$, $n \le 2$, for at least some field orientations.

PrOs$_4$Sb$_{12}$ is a superconductor ($T_c = 1.85$~K~\cite{BFHZ02}) with a number of extraordinary properties~\cite{MHZY03}. It is the only known Pr-based heavy-fermion superconductor, the Pr$^{3+}$ ground state is nonmagnetic, a novel ordered phase appears at high fields and low temperatures, there are multiple superconducting phases, and TRS is broken in the superconducting state~\cite{ATKS03}. Previous $\mu$SR measurements of $\lambda$ in the vortex state of a powdered sample~\cite{MSHB02} found evidence for a BCS-like activated dependence at low temperatures, suggesting the absence of gap nodes. But radiofrequency (rf) inductive measurements of the surface penetration depth~\cite{CSSS03} indicate point nodes in the gap, in disagreement with the $\mu$SR results. 

This Letter reports new $\mu$SR experiments on oriented PrOs$_4$Sb$_{12}$ crystals, and compares $\mu$SR and surface penetration-depth measurements in PrOs$_4$Sb$_{12}$ and a number of other superconductors. The discrepancy between these measurements noted above is found in TRS-breaking superconductors but not otherwise, and is therefore correlated with TRS breaking.

In the vortex lattice of a type-II superconductor each vortex possesses a normal-state-like core surrounded by a shielding supercurrent. These supercurrents give rise to an inhomogeneous magnetic field that is periodic in the vortex lattice. The vortex-lattice field distribution~$P_{\text{v}}(B)$ is related to $\lambda$. For a fixed density of vortices the rms width~$\delta B$ of $P_{\text{v}}(B)$ decreases with increasing $\lambda$; in a London superconductor ($\lambda$ much longer than the coherence length~$\xi$) $\delta B \propto 1/\lambda^2$ when the applied field~$H$ satisfies $H_{c1} \ll H \ll H_{c2}$~\cite{Bran88}. Transverse-field $\mu$SR (TF-$\mu$SR) experiments (field applied transverse to the $\mu^+$ spin) are sensitive probes of this field distribution~\cite{SBK00}.

Time-differential TF-$\mu$SR experiments were carried out at the M15 channel at \mbox{TRIUMF}, Vancouver, Canada, on a mosaic of oriented PrOs$_4$Sb$_{12}$ crystals. The crystals were mounted on a thin GaAs backing, which rapidly depolarizes muons in transverse field and minimizes any spurious signal from muons that do not stop in the sample. $\mu$SR asymmetry data were taken for temperatures in the range 0.02--2.5~K and $\mu_0H$ between 10~mT and 100~mT applied parallel to the $\langle100\rangle$ axes of the crystals. The data were fitted with the functional form~$G(t)\cos(\omega t + \phi)$, where the frequency~$\omega$ and phase~$\phi$ describe the average $\mu^+$ precession and the relaxation function~$G(t)$ describes the loss of phase coherence due to the distribution of precession frequencies~\cite{static}. The relaxation rate associated with $G(t)$ is a measure of the width of this distribution and hence of $\delta B$~\cite{SBK00}.

Neither of the commonly-used exponential or Gaussian functional forms accurately fit the asymmetry data in the normal state, due to $\mu^+$ coupling to both nuclear and Pr$^{3+}$ spins as discussed below. Data from both the normal and the superconducting states are well fit, however, by either of two slightly more complex functional forms: the ``power exponential''
\begin{equation}
G(t) = \exp[-(\Lambda t)^K ]\,,
\label{eq:pe}
\end{equation} 
and the damped Gaussian
\begin{equation}
G(t) = e^{-Wt} \exp(-{\textstyle\frac{1}{2}} \sigma^2t^2) \,.
\label{eq:dg}
\end{equation}
These functions are both phenomenological and have no theoretical motivation. We shall see that the su\-per\-con\-ducting-state properties obtained from these fits are similar for both functions, which indicates insensitivity to details of the fitting function and justifies {\em a posteriori\/} these otherwise arbitrary choices.

\begin{figure}[ht]
\includegraphics[clip=,width=0.86\columnwidth]{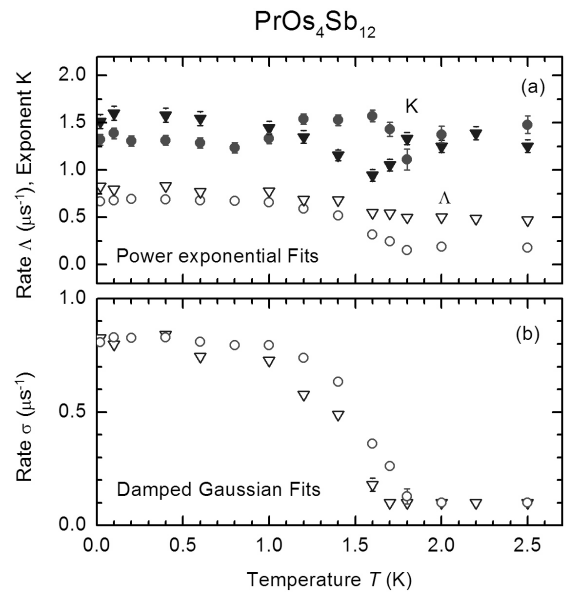}
\caption{Relaxation rates from TF-$\mu$SR asymmetry data in PrOs$_4$Sb$_{12}$ for applied fields~$\mu_0H = 10$~mT (circles) and 100~mT (triangles). (a)~Relaxation rates~$\Lambda$ (open symbols) and exponents~$K$ (filled symbols) from power-exponential fits [Eq.~(\protect\ref{eq:pe})]. Note that $\Lambda$ and $K$ have different dimensions; they are plotted on the same graph only for convenience. (b)~Gaussian relaxation rates~$\sigma$ from damped Gaussian fits [Eq.~(\protect\ref{eq:dg})]. The exponential rates have been fixed to the normal-state values [$W(10~\text{mT}) = 0.095\ \mu\text{s}^{-1}$, $W(100~\text{mT}) = 0.428\ \mu\text{s}^{-1}$].}
\label{fig:relax}
\end{figure}

Figure~\ref{fig:relax}(a) shows the temperature dependence of the relaxation rate~$\Lambda$ and exponent~$K$ from power-ex\-po\-nent\-ial fits for $\mu_0H = 10$~mT and 100~mT\@. 
The relaxation is intermediate between exponential and Gaussian ($1 \lesssim K < 2$). The temperature dependence of $K$ below $T_c$ indicates that the shape of $G(t)$ is changing with temperature, but for temperatures lower than about 1~K $\Lambda$ and $K$ are nearly constant for $\mu_0H =10$~mT. 

Figure~\ref{fig:relax}(b) gives the temperature dependence of the Gaussian rate~$\sigma$ from damped-Gaussian fits. Below $T_c$ the exponential rate~$W$ [Eq.~(\ref{eq:dg})] was held fixed at the normal-state value for each field, so that the temperature dependence of the Gaussian rate~$\sigma$ reflects the effect of the superconducting state. Some such procedure is necessitated by the strong statistical correlation between $W$ and $\sigma$ in Eq.~(\ref{eq:dg}); the time constant and the shape of the function are influenced by both of these parameters, so that correlations between them can result from small systematic errors. The principal justification for this {\em ad hoc\/} fixing of $W$ is the insensitivity of the superconducting-state results to details of the fitting function noted above. Statistical correlation is less of a problem in power exponential fits, where $\Lambda$ and $K$ control the rate and shape, respectively, and are not as strongly correlated.

For damped-Gaussian fits the normal-state Gaussian rate is independent of field [cf.\ Fig.~\ref{fig:relax}(b)] and of magnitude consistent with nuclear dipolar broadening. The exponential rate increases with field and is due to inhomogeneity in the Pr$^{3+}$ magnetization. Determination of the vortex-state field distribution width requires correction for the normal-state relaxation. For the damped Gaussian fits we take the superconducting-state Gaussian rate~$\sigma_s$ to be given by $\sigma_s^2 = \sigma^2 - \sigma_n^2$, where $\sigma_n$ is the normal-state rate~\cite{SBK00}. For the power exponential fits the proper correction procedure is somewhat less clear. We have chosen to use the relation $\Lambda_s^K = \Lambda^K - \Lambda_n^K$, which interpolates smoothly between the exponential ($K = 1$) and Gaussian ($K = 2$) limits. The results are sensitive to this choice only for $\mu_0H = 100$~mT, where $\Lambda_n$ is significant [cf.\ Fig.~\ref{fig:relax}(a)].

\begin{figure}[ht]
\includegraphics[clip=,width=0.86\columnwidth]{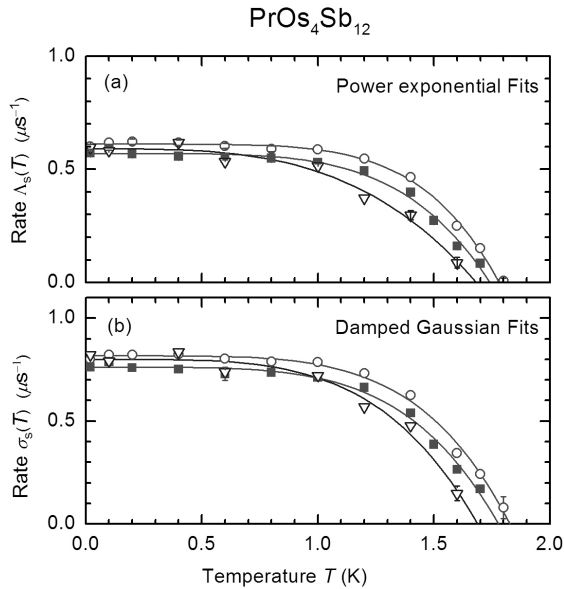}
\caption{Temperature dependence of superconducting-state relaxation rates in PrOs$_4$Sb$_{12}$, corrected for normal-state relaxation (see text). (a)~power exponential rates~$\Lambda_s$. (b)~damped Gaussian rates~$\sigma_s$. Circles: applied field~$\mu_0H = 10$ mT\@. Squares: $\mu_0H = 20$ mT\@. Triangles: $\mu_0H = 100$ mT\@. The curves are guides to the eye.}
\label{fig:scstate}
\end{figure}

Figure~\ref{fig:scstate} gives the temperature dependence of the corrected superconducting-state $\mu^+$ relaxation rates for $\mu_0H = 10$, 20, and 100~mT\@. 
It can be seen that the qualitative behavior of the rates is remarkably independent of the fit function. At 10 and 20~mT both $\Lambda_s$ and $\sigma_s$ are nearly temperature-independent below $\sim$1~K\@. At the lowest temperatures the rate is field-in\-de\-pen\-dent to within a few percent between 10 mT and 100 mT\@. In an isotropic superconductor such as cubic PrOs$_4$Sb$_{12}$ vortex-lattice disorder is expected to increase the low-field rate; increasing field (increasing vortex density) then decreases the rate as intervortex interactions stabilize the lattice~\cite{SBK00,NBHK02}. Thus the field independence of the rate indicates a substantially ordered vortex lattice, in which case the temperature dependence of the rate is controlled solely by the temperature dependence of the field distribution~$P_{\text{v}}(B)$. We can also conclude that the field dependence expected as $H \to H_{c1}$~\cite{SSF90} plays no role.

The expression~\cite{Bran88}
\begin{equation}
\delta B^2(T) = 0.00371\, \Phi_0^2\lambda^{-4}(T) \,,
\label{eq:Brandt}
\end{equation}
where $\Phi_0$ is the flux quantum, relates the second moment~$\delta B^2$ of $P_{\text{v}}(B)$ to $\lambda$ for a triangular vortex lattice in the London limit. The second moment of the corresponding $\mu^+$ frequency distribution is $\delta\omega^2 = \gamma_\mu^2\delta B^2$, where $\gamma_\mu$ is the $\mu^+$ gyromagnetic ratio. Then the $\mu$SR estimate~$\lambda_{\mu\text{SR}}$ of the penetration depth from Eq.~(\ref{eq:Brandt}) is
\begin{equation}
\lambda_{\mu\text{SR}}\ (\mu\text{m}) = 0.328/\sqrt{\delta\omega\ (\mu\text{s}^{-1})} \,.
\label{eq:lambda}
\end{equation}
Now the rms width~$\sigma_s$ of the best-fit Gaussian is not necessarily $\delta\omega$, and replacement of $\delta\omega$ in Eq.~(\ref{eq:lambda}) by $\sigma_s$ is not completely justified. Nevertheless $\sigma_s$ should scale with $\delta\omega$, and within its range of validity Eq.~(\ref{eq:lambda}) should give the correct temperature dependence of $\lambda_{\mu\text{SR}}$. This is because under these circumstances effects of nonzero $\xi$ are restricted to the high-field tail of $P_{\text{v}}(B)$, which is not heavily weighted in a Gaussian fit (cf.\ Fig.~1 of Ref.~\cite{Kado04}). PrOs$_4$Sb$_{12}$ is a strongly type-II superconductor (Ginzburg-Landau $\kappa = \lambda/\xi \approx 30$~\cite{BFHZ02,MSHB02}), and this picture should be applicable.

\begin{figure}[ht]
\includegraphics[clip=,width=0.86\columnwidth]{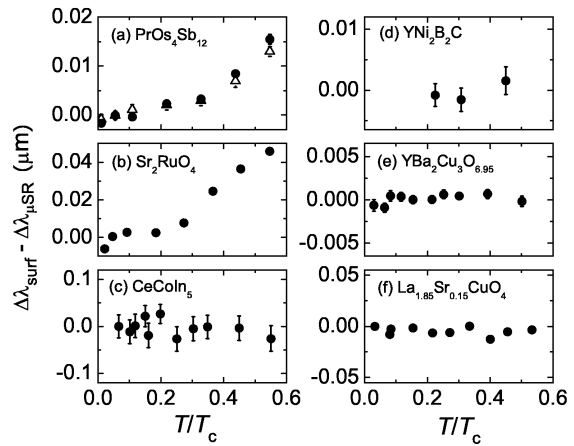}
\caption{Dependence of difference~$\Delta\lambda_{\text{surf}} - \Delta\lambda_{\mu\text{SR}}$ (see text) on reduced temperature $T/T_c$ in six superconductors. (a) PrOs$_4$Sb$_{12}$. 
 Triangles: $\lambda_{\mu\text{SR}}$ from $\Lambda_s$ [Fig.\ \protect\ref{fig:scstate}(a)]. Circles: $\lambda_{\mu\text{SR}}$ from $\sigma_s$ [Fig.\ \protect\ref{fig:scstate}(b)]. (b) Sr$_2$RuO$_4$. 
(c) CeCoIn$_5$. 
(d) YNi$_2$B$_2$C. 
(e) YBa$_2$Cu$_3$O$_{6.95}$. 
(f) La$_{1.85}$Sr$_{0.15}$CuO$_4$. 
See text for references.}
\label{fig:depth}
\end{figure}

Figure~\ref{fig:depth}(a) compares $\Delta\lambda_{\text{surf}}(T) = \lambda_{\text{surf}}(T) - \lambda_{\text{surf}}(0)$, obtained from rf inductance measurements in the Meissner state~\cite{CSSS03,deltalambda}, with $\Delta\lambda_{\mu\text{SR}}(T)$ obtained using $\mu^+$ relaxation rates for $\mu_0H = 10$~mT (Fig.~\ref{fig:scstate}) in Eq.~(\ref{eq:lambda}). 
At low temperatures~\cite{lowtemp} the difference~$\Delta\lambda_{\text{surf}}(T) - \Delta\lambda_{\mu\text{SR}}(T)$ increases markedly with increasing temperature; this is the discrepancy between the measurements noted above. It is the same whether $\Lambda_s(T)$ [Fig.~\ref{fig:scstate}(a)] or $\sigma_s(T)$ [Fig.~\ref{fig:scstate}(b)] is used in Eq.~(\ref{eq:lambda}); the comparison does not depend on the choice of fitting function. Figure~\ref{fig:depth}(b) gives $\Delta\lambda_{\text{surf}}(T) - \Delta\lambda_{\mu\text{SR}}(T)$ for the TRS-breaking superconductor~Sr$_2$RuO$_4$, using data from the literature~\cite{LFKL00,BYSVH00}. Again there is a discrepancy (as noted previously~\cite{MaMa03}), which is very similar to that in PrOs$_4$Sb$_{12}$. Small-angle neutron diffraction experiments~\cite{KRFG00} also found a temperature-independent vortex-lattice field distribution in Sr$_2$RuO$_4$ at low temperatures. 

Figures~\ref{fig:depth}(c)--(f) give $\Delta\lambda_{\text{surf}}(T) - \Delta\lambda_{\mu\text{SR}}(T)$ from literature data for the heavy-fermion compound~CeCoIn$_5$~\cite{HKKK02,CVSY03}, the borocarbide YNi$_2$B$_2$C~\cite{CHBC94,STYC99}, and the high-$T_c$ cuprates~YBa$_2$Cu$_3$O$_{6.95}$~\cite{SBKM99,HBML93} and La$_{1.85}$Sr$_{0.15}$CuO$_4$~\cite{LFKL97,PRCL99}. None of these superconductors exhibit TRS breaking, and none exhibit the temperature dependence of $\Delta\lambda_{\text{surf}} - \Delta\lambda_{\mu\text{SR}}$ seen in Figs.~\ref{fig:depth}(a) and (b). {\em There is a significant difference between\/ \rm $\Delta\lambda_{\text{surf}}(T)$ \em and\/ \rm $\Delta\lambda_{\mu\text{SR}}(T)$ \em only for the TRS-breaking superconductors~PrOs$_4$Sb$_{12}$ and Sr$_2$RuO$_4$.\/} This is in spite of a wide range of experimental methods and analysis techniques: in Figs.~3(a), (b), (c), and (f) $\Delta\lambda_{\text{surf}}$ was obtained from rf measurements, in (d) from magnetization, and in (e) from microwave impedance; in Figs.~3(a) and (c) $\lambda_{\mu\text{SR}}$ was obtained from Eq.~(\ref{eq:lambda}), in (b), (e), and (f) from fits to the expected vortex-state field distribution, and in (d) from a generalization of Eq.~(\ref{eq:lambda}) to low and high fields~\cite{CHBC94}.

The origin of this discrepancy and its relation to TRS breaking are not clear. Low-field low-temperature phase transitions between superconducting states have been reported in both PrOs$_4$Sb$_{12}$~\cite{CMSF04} and Sr$_2$RuO$_4$~\cite{MDAM99}, and may be involved in the discrepancy. The TRS-breaking state may couple to rf or microwave fields, necessitating a revised interpretation of the surface measurements. A mechanism of this sort, in which subgap chiral surface states affect the surface penetration depth, has been proposed for Sr$_2$RuO$_4$~\cite{MoSi00}. A $T^2$ power law is found for surface measurements even though the bulk energy spectrum is gapped. The theory requires $\lambda \approx \xi$, however, and thus seems inapplicable to PrOs$_4$Sb$_{12}$. It has also been noted~\cite{YiGa93} that surface scattering breaks pairs in an odd-parity superconductor. To our knowledge the surface penetration depth has not been calculated taking this effect into account, but pair breaking would decrease the gap and therefore increase the temperature dependence of $\lambda$. The discrepancy might be related to a breakdown of the relation~$\delta B(T) \propto 1/\lambda^2$ due to nonlinear/nonlocal effects~\cite{SBK00,AFA00}, or to the spontaneous magnetic field in the vortex state~\cite{ATKS03} (although the measured field is not large enough to have a significant direct effect on $\delta B$).

Sr$_2$RuO$_4$ and PrOs$_4$Sb$_{12}$ are both TRS-breaking superconductors but are otherwise very different. Tetragonal Sr$_2$RuO$_4$ is an anisotropic transition-metal-oxide superconductor that is weakly type-II for $\mathbf{H \parallel c}$ ($\kappa_{ab} = 2.3$~\cite{MaMa03}), whereas cubic PrOs$_4$Sb$_{12}$ is an isotropic strongly type-II heavy-fermion superconductor. The fact that a similar discrepancy between $\mu$SR and surface measurements of $\lambda(T)$ is found in such different materials, but not in a variety of non-TRS-breaking superconductors, strongly suggests that TRS breaking is involved.

We are grateful for discussions with I. Affleck, E.~D. Bauer, K. Maki, D. Parker, L.~P. Pryadko, and C.~M. Varma, to the \mbox{TRIUMF} staff for help during the experiments, and to S.~K. Kim for help in in sample preparation. This work was supported in part by the U.S. NSF, Grants DMR-0102293 (Riverside), DMR-0203524 (Los Angeles), and DMR-0335173 (San Diego), by the Canadian NSERC and CIAR (Burnaby), and by the U.S. DOE (San Diego, Grant DE-FG02-04ER46105). Work at Los Alamos was performed under the auspices of the U.S. DOE.



\begin{thebibliography}{10}

\bibitem{SiUe91}
M. Sigrist and K. Ueda, Rev. Mod. Phys. {\bf 63},  239  (1991).

\bibitem{HeNo96}
R.~H. Heffner and M.~R. Norman, Comments Condens. Matter Phys. {\bf 17},  361
  (1996).

\bibitem{SBK00}
J.~E. Sonier, J.~H. Brewer, and R.~F. Kiefl, Rev. Mod. Phys. {\bf 72},  769
  (2000).

\bibitem{Brew94}
J.~H. Brewer,  in {\em Encyclopedia of Applied Physics}, edited by G.~L. Trigg
  (VCH Publishers, New York, 1994), Vol.~11, p.\ 23.

\bibitem{HSWB90}
R.~H. Heffner {\it et~al.}, Phys. Rev. Lett. {\bf 65},  2816  (1990).

\bibitem{LFKL98}
G.~M. Luke {\it et~al.}, Nature {\bf 394},  558  (1998).

\bibitem{ATKS03}
Y. Aoki {\it et~al.}, Phys. Rev. Lett. {\bf 91},  067003  (2003).

\bibitem{BFHZ02}
E.~D. Bauer {\it et~al.}, Phys. Rev. B {\bf 65},  100506(R)  (2002).

\bibitem{MHZY03}
M.~B. Maple {\it et~al.}, Physica C {\bf 388-389},  549  (2003).

\bibitem{MSHB02}
D.~E. MacLaughlin {\it et~al.}, Phys. Rev. Lett. {\bf 89},  157001  (2002).

\bibitem{CSSS03}
E.~E.~M. Chia, M.~B. Salamon, H. Sugawara, and H. Sato, Phys. Rev. Lett. {\bf
  91},  247003  (2003).

\bibitem{Bran88}
E.~H. Brandt, Phys. Rev. B {\bf 37},  R2349  (1988).

\bibitem{static}
Zero- and longitudinal-field $\mu$SR experiments \protect\cite{ATKS03} show
  that dynamic (homogeneous) relaxation rates are negligible at the applied
  fields used in this work.

\bibitem{NBHK02}
Ch. Niedermayer {\it et~al.}, Phys. Rev. B {\bf 65},  094512  (2002).

\bibitem{SSF90}
A.~D. Sidorenko, V.~P. Smilga, and V.~I. Fesenko, Hyperfine Interact. {\bf 63},
   49  (1990).

\bibitem{Kado04}
R. Kadono, J. Phys.: Condens. Matter {\bf 16},  S4421  (2004).

\bibitem{deltalambda}
We compare $\Delta\lambda(T)$ rather than $\lambda(T)$ because the inductive
  technique does not give an accurate experimental value of $\lambda(0)$.

\bibitem{lowtemp}
The superconducting gap structure (presence or absence of nodes) is reflected
  in $\lambda(T)$ only at low temperatures.

\bibitem{LFKL00}
G.~M. Luke {\it et~al.}, Physica B {\bf 289-290},  373  (2000).

\bibitem{BYSVH00}
I. Bonalde {\it et~al.}, Phys. Rev. Lett. {\bf 85},  4775  (2000).

\bibitem{MaMa03}
A.~P. Mackenzie and Y. Maeno, Rev. Mod. Phys. {\bf 75},  657  (2003).

\bibitem{KRFG00}
P.~G. Kealey {\it et~al.}, Phys. Rev. Lett. {\bf 84},  6094  (2000).

\bibitem{HKKK02}
W. Higemoto {\it et~al.}, J. Phys. Soc. Jpn. {\bf 71},  1023  (2002).

\bibitem{CVSY03}
E.~E.~M. Chia {\it et~al.}, Phys. Rev. B {\bf 67},  014527  (2003).

\bibitem{CHBC94}
R. Cywinski {\it et~al.}, Physica C {\bf 233},  273  (1994).

\bibitem{STYC99}
K.~J. Song {\it et~al.}, Phys. Rev. B {\bf 59},  R6620  (1999).

\bibitem{SBKM99}
J.~E. Sonier {\it et~al.}, Phys. Rev. Lett. {\bf 83},  4156  (1999).

\bibitem{HBML93}
W.~N. Hardy {\it et~al.}, Phys. Rev. Lett. {\bf 70},  3999  (1993).

\bibitem{LFKL97}
G.~M. Luke {\it et~al.}, Physica C {\bf 282-287},  1465  (1997).

\bibitem{PRCL99}
C. Panagopoulos {\it et~al.}, Phys. Rev. B {\bf 60},  14617  (1999).

\bibitem{CMSF04}
T. Cichorek {\it et~al.}, cond-mat/0409331 (unpublished).

\bibitem{MDAM99}
A.~C. Mota, E. Dumont, A. Amann, and Y. Maeno, Physica B {\bf 259-261},  934
  (1999).

\bibitem{MoSi00}
T. Morinari and M. Sigrist, J. Phys. Soc. Jpn. {\bf 69},  2411  (2000).

\bibitem{YiGa93}
S. Yip and A. Garg, Phys. Rev. B {\bf 48},  3304  (1993).

\bibitem{AFA00}
M.~H.~S. Amin, M. Franz, and I. Affleck, Phys. Rev. Lett. {\bf 84},  5864
  (2000).

\end{thebibliography}

\end{document}